\documentclass[10pt,twocolumn,letterpaper]{article}

\usepackage{cvpr}              % To produce the CAMERA-READY version
\usepackage[dvipsnames]{xcolor}
\newcommand{\red}[1]{{\color{red}#1}}

\usepackage{graphicx}
\usepackage{amsmath}
\usepackage{amssymb}
\usepackage{booktabs}
\usepackage{multirow}
\usepackage{multicol}
\usepackage{marvosym}

\usepackage[export]{adjustbox}

\definecolor{cvprblue}{rgb}{0.21,0.49,0.74}
\usepackage[pagebackref,breaklinks,colorlinks,citecolor=cvprblue]{hyperref}

\usepackage[capitalize]{cleveref}
\crefname{section}{Sec.}{Secs.}
\Crefname{section}{Section}{Sections}
\Crefname{table}{Table}{Tables}
\crefname{table}{Tab.}{Tabs.}

\def\confName{CVPR}
\def\confYear{2024}

\author{Cheng Wan\\
Institution1\\
Institution1 address\\
{\tt\small cwan38@gatech.edu}
\and
Second Author\\
Institution2\\
First line of institution2 address\\
{\tt\small secondauthor@i2.org}
}

\begin{document}

\title{Swift Parameter-free Attention Network for Efficient Super-Resolution}
\author{Cheng Wan$^{1}$\thanks{Equal contribution, \textsuperscript{\Letter}Corresponding Author} \quad Hongyuan Yu$^{2*}$\textsuperscript{\Letter}  \quad Zhiqi Li$^{1*}$ \quad Yihang Chen$^{1}$   \quad Yajun Zou$^{2}$  \\
\quad Yuqing Liu$^{2}$ \quad Xuanwu Yin$^{2}$ \quad Kunlong Zuo$^{2}$\\ %\thanks{Corresponding author}
$^{1}$Georgia Institute of Technology, 
$^{2}$Xiaomi Inc \\
{\tt\small \textcolor{magenta}{\{cwan38, zli3167, ychen3726\}@gatech.edu}}\\
{\tt\small \textcolor{magenta}{\{yuhongyuan, zouyajun, liuyuqing9, yinxuanwu, zuokunlong\}@xiaomi.com}}\\
\vspace{-2em}
}

\maketitle
% \vspace{-2.5em}
\vspace{-1.5em} 
\begin{abstract}
Single Image Super-Resolution (SISR) is a crucial task in low-level computer vision, aiming to reconstruct high-resolution images from low-resolution counterparts. Conventional attention mechanisms have significantly improved SISR performance but often result in complex network structures and large number of parameters, leading to slow inference speed and large model size. To address this issue, we propose the Swift Parameter-free Attention Network (SPAN), a highly efficient SISR model that balances parameter count, inference speed, and image quality. SPAN employs a novel parameter-free attention mechanism, which leverages symmetric activation functions and residual connections to enhance high-contribution information and suppress redundant information. Our theoretical analysis demonstrates the effectiveness of this design in achieving the attention mechanism's purpose. We evaluate SPAN on multiple benchmarks, showing that it outperforms existing efficient super-resolution models in terms of both image quality and inference speed, achieving a significant quality-speed trade-off. This makes SPAN highly suitable for real-world applications, particularly in resource-constrained scenarios. Notably, we won the first place both in the overall performance track and runtime track of the NTIRE 2024 efficient super-resolution challenge. Our code and models are made publicly available at \url{https://github.com/hongyuanyu/span}.
\vspace{-1em}
\end{abstract}

%%%%%%%%% BODY TEXT
\vspace{-1em}
\section{Introduction}
\vspace{-1em}
\label{sec:intro}

Single Image Super-Resolution (SISR) is a well-established task in low-level computer vision, which aims to reconstruct a high-resolution image from a single low-resolution image. This task has broad applicability in enhancing image quality across various domains \cite{singh2016super,sreenivas2011processing,genitha2010super,liu2005hallucinating,yuan2009fingerprint,peeters2004use,patel2012super}. The advent of deep learning has led to significant advancements in this field \cite{SRCNN, FSRCNN, VDSR, CARN, IMDN, RFDN, RFANet, ECBSR, SwinIR, sun2022shufflemixer}. Recent progress in super-resolution tasks has been largely driven by the attention mechanism.  Numerous state-of-the-art super-resolution networks incorporate attention mechanisms or even employ larger vision transformers (ViTs) as the model architecture~\cite{kong2022residual, wang2023omni, liu2020residual, hui2019lightweight, cao2023ciaosr, MAFFSRN, RCAN, SwinIR, choi2023n}. These networks emphasize key features and long-distance dependencies between patches through attention maps, capturing a wider range of contextual information to ensure continuity of details and accuracy of edge textures. However, the computational requirements of the attention mechanism, which involve complex network structures and a substantial number of additional parameters, lead to challenges such as large model size and slow inference speed. These challenges limit the applicability of these models, hindering their use in efficient, high-speed computing scenarios, such as SISR tasks on resource-constrained mobile devices.

\begin{figure}[!t]
\vspace{-1.5em} 
\includegraphics[width=\linewidth]{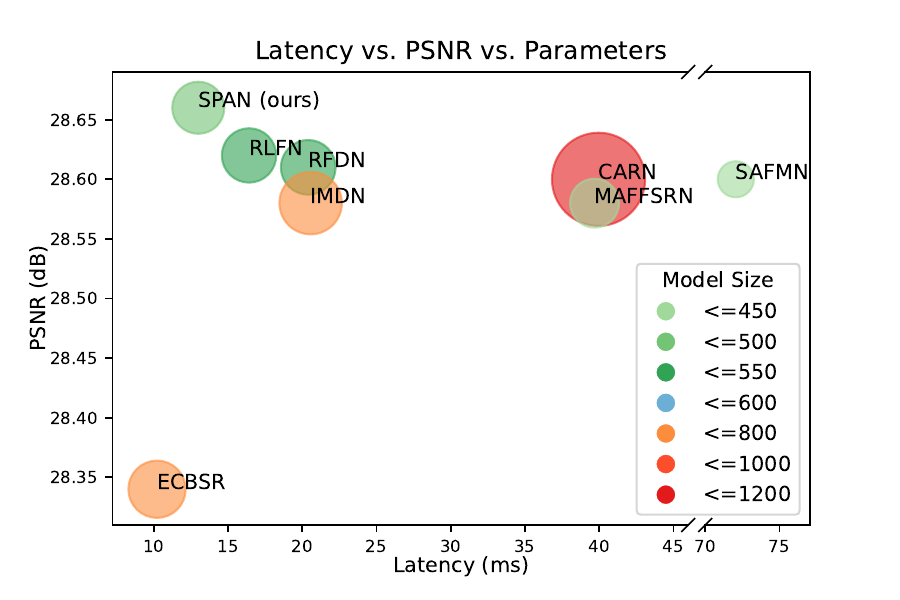}
\caption{Latency, PSNR, and complexity of model comparison on Set14 dataset in x4 scale factor task.}
\label{fig:comparison}
\vspace{-1.5em} 
\end{figure}

Numerous existing efficient super-resolution (ESR) techniques have achieved certain successes in enhancing model efficiency. Some models primarily focus on reducing model FLOPs and parameters, accomplishing this through methods like group convolution and depth-wise separable convolution~\cite{CARN, IMDN, RFDN}. 
However, simply reducing FLOPs or parameters sometimes does not lead to a significant improvement in the model's inference speed, and it can also decrease model accuracy. Other models~\cite{DRCN} reduce model parameter size through feature information sharing and downsizing non-attention branches. But these models still contain many parameters within their complex computational structures, resulting in long running time. 
To ensure fast inference speed, it is crucial to maintain a simple network topology. However, conventional attention mechanisms often result in more complex network structures. To address this problem, we propose a parameter-free attention mechanism and theoretically demonstrate that our Swift Parameter-free Attention Network (SPAN) can achieve the attention mechanism's role of enhancing high-contribution information and suppressing redundant information through symmetric activation functions and residual connections.

In SPAN, a parameter-free attention mechanism is constructed by passing extracted features through a symmetric activation function around the origin to calculate attention maps directly. This attention mechanism focuses on information-rich regions without the need of additional parameter learning, allowing for rapid and effective feature extraction from shallow to deep layers. The design of symmetric activation functions and residual connections in the modules help to solve issues related to information loss of the parameter-free attention modules. The simplicity of the network structure ensures operational speed, addressing the challenges posed by conventional attention mechanisms.

In summary, our main contributions are as follows:

\begin{itemize}
    \item We design a novel parameter-free attention mechanism that employs symmetric activation functions and residual connections to enhance high-contribution information and suppress redundant information, thereby simplifying the network structure and improving inference speed without sacrificing accuracy.
    \item We propose the Swift Parameter-free Attention Network (SPAN), which leverages the parameter-free attention mechanism to achieve rapid and effective feature extraction from shallow to deep layers while maintaining low model complexity and parameter count.
    \item Through theoretical analysis and experimental validation, we demonstrate the effectiveness and superiority of SPAN in single-image super-resolution tasks, proving its practicality and potential application value in resource-constrained scenarios.
\end{itemize}

\vspace{-0.5em} 
\section{Related Work}
\vspace{-0.5em} 
\subsection{Efficient Super Resolution on Image}
\vspace{-0.5em} 
Most existing ESR models focus on reducing model parameters or FLOPs to improve efficiency. SRCNN~\cite{dong2014learning} introduces an end-to-end mapping for single image super-resolution using a deep convolutional neural network (CNN), optimizing all layers jointly. DRCN~\cite{kim2016deeply} presents a deeply-recursive convolutional network with up to 16 recursive layers to enhance super-resolution performance, overcoming training challenges through innovative techniques and achieving substantial improvements over previous methods. LatticeNet~\cite{luo2020latticenet} introduces the Lattice Block for combining Residual Blocks using a lattice filter bank, showcasing improved performance through this novel combination approach. CARN~\cite{ahn2018fast} implements a cascading mechanism on a residual network to create an accurate and lightweight model. IMDN~\cite{hui2019lightweight} is a lightweight and accurate single image SR model. It extracts hierarchical features and selectively aggregates them using contrast-aware channel attention. RFDN~\cite{liu2020residual} improves over IMDN using more lightweight and flexible feature distillation connections and shallow residual blocks, achieving better SR performance with lower model complexity. However, simply minimizing parameters and FLOPs does not necessarily lead to better model efficiency, especially during inference. There is a need to develop SR models that prioritize faster inference speed rather than just reducing parameters or FLOPs.

To address this, RLFN~\cite{kong2022residual} enhances model compactness and accelerates inference without sacrificing SR restoration quality based on RFDN. They analyze properties of intermediate features and find shallow features are critical for PSNR-oriented models. Based on this, they propose an improved feature extractor to effectively capture edges and details. Further, a multi-stage warm-up training strategy is introduced to speed up model convergence and improve SR restoration accuracy. Omni~\cite{wang2023omni} introduces Aggregation Networks for efficient lightweight image super-resolution. It utilizes Omni Self-Attention to fuse spatial and channel self-attentions. In addition, a multi-scale feature extraction method is introduced to achieve high-quality restoration with low computational cost. 

\begin{figure*}[ht]
\vspace{-0.5em} 
\centering
\includegraphics[width=0.9\linewidth]{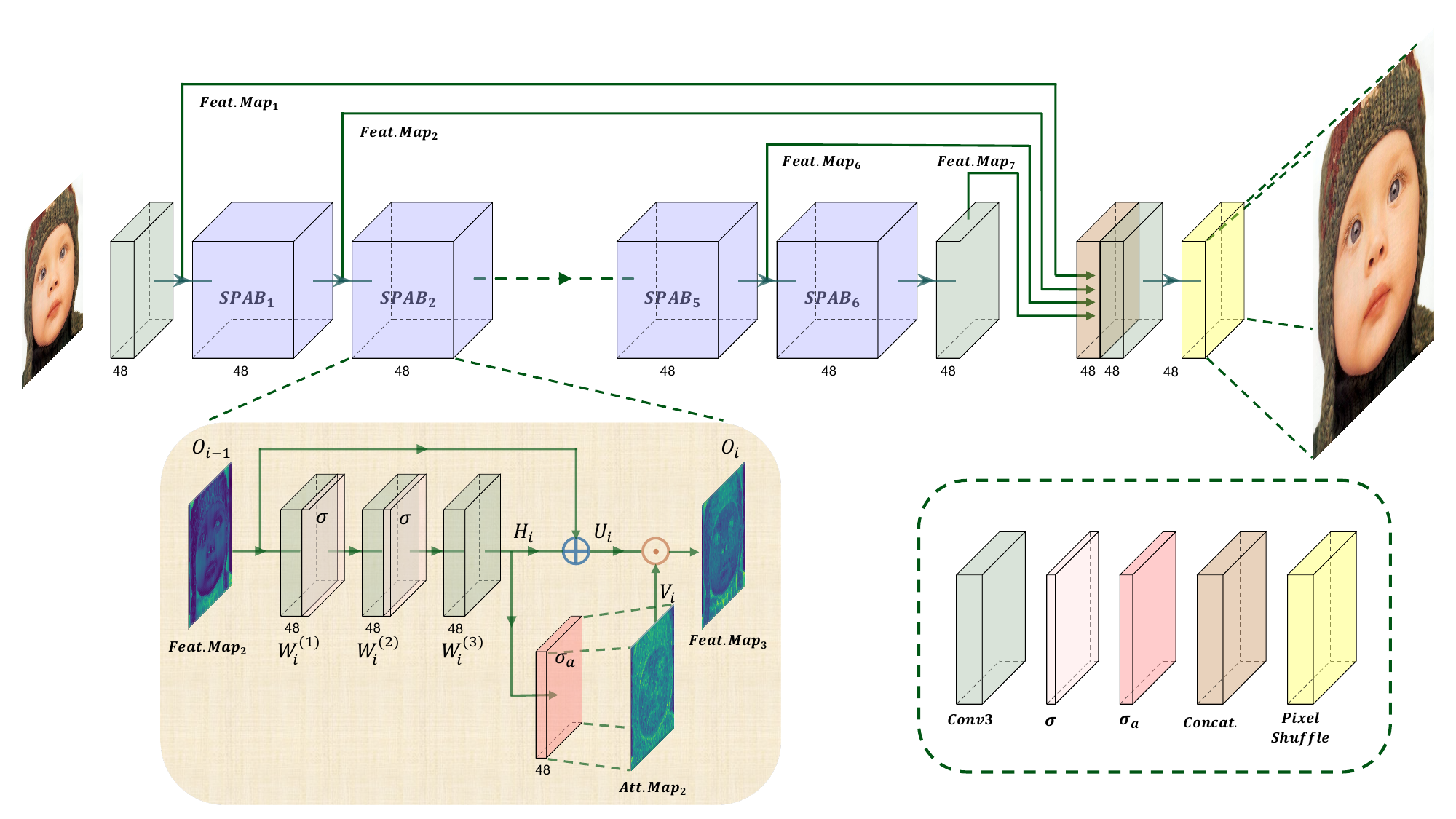}
\vspace{-0.5em} 
\caption{The proposed SPAN architecture. The yellow area indicates the internal structure of each SPAB module. $Att. Map_2$ denotes the generated attention map. Input is a low resolution image, and output is a high resolution image.} %  The dotted line indicates that there are two more SPAB modules directly connected in series, and their output feature maps are not needed to participate in concatenate operation.
\label{fig:pipeline}
\vspace{-1.5em} 
\end{figure*}

\vspace{-0.5em} 
\subsection{Attention Mechanism}
\vspace{-0.5em} 

For ESR task, the application of lightweight attention mechanisms play a significant role for enhancing model performance without substantially increasing complexity. The pivotal role of attention in modern vision models lies in its dynamic re-weighting of features, which directs computational resources to the most salient parts of the input, thus boosting efficiency and efficacy across various tasks~\cite{yu2022memory, truong2022direcformer, lee2022knn, jang2022glamd, cao2022learning, gao2022aiatrack}.

Attention-based super-resolution networks typically require a substantial receptive field to capture both local and global information, thereby enhancing super-resolution performance. However, the utilization of parameterized attention maps can slow down inference speed. In contrast, efficient super-resolution (SR) networks should maintain performance while ensuring rapid inference speed.

We observe that attention maps can be generated without necessitating additional training and parameters~\cite{choe2019attention, haase2020rethinking}, yet still contribute positively to the model's performance. The crux of this lightweight attention approach is to maximize the representational power of the super-resolution network within a constrained model budget. By incorporating these principles, we can develop a swift and effective attention mechanism for super-resolution models.
Our proposed parameter-free attention mechanism utilizes existing convolutional layers to enhance high-contribution information and suppress redundant information, thus eliminating the need for additional parameter-intensive processes. This not only streamlines the model but also augments its ability to weakly localize objects, a feature that is essential for refining super-resolution techniques~\cite{choe2019attention}.

\vspace{-0.5em}
\section{Method}
\label{sec:model-arch}

%图中对SPAN，没有残差连接的SPAN，和没有attention也没有残差连接的SPAN的第一个和第六个SPAB输出的feature map进行了对比，可以发现只有attention没有残差连接的SPAB会丢失过多的信息，加上残差连接过后可以缓解这个问题

%从左到右依次是激活函数$\sigma_a$为sigmoid时第一个SPAB的Pre-attention Feature Map $H_1$，Attention Map $V_1$和Output Feature Map $O_i$，亮度大的区域表示绝对值大。对比$H_1$和$V_1$可以发现，在$H_1$中特征绝对值大的区域，在$V_1=\sigma_a(H_i)$中反而由于sigmoid滤掉负数的作用而变小，表现的比较暗，通过使用这样的attention map $V_1$得到的最终feature map $O_i$中特征没有被突出，反而特征更加模糊。

%在这一节中，我们将首先展示我们提出的方法：基于自注意力机制的Swift Parameter-free Attention Block(SPAB)和应用于Super resolution的基于SPAB搭建的网络Swift Parameter-free Attention Network（SPAN），然后对SPAB进行理论分析，论证了在残差连接和关于原点对称的激活函数的作用下，对于super resolution 的任务，parameter free的self-attention机制的有效性。 
In this section, we will first present our proposed method: the Swift Parameter-free Attention Block (SPAB) based on the attention mechanism and the proposed SPAN built using SPABs for super resolution. Then, we will conduct theoretical analysis of SPAB, demonstrating the effectiveness of parameter-free attention mechanisms for super resolution tasks, especially with symmetric activation functions and residual connections.

 \vspace{-0.5em}
\subsection{Network Architecture.}
\label{sec:architechture}

%如图片\autoref{fig:pipeline}所示，SPAN包含了六个连续的SPAB块，每个SPAB块通过三层$H' \times W'$的卷积层提取高频特征，提取到的特征和输入进行残差连接过后作为该层的feature map $U_i$, 而卷积层提取的特征通过激活函数$\sigma(\cdot)$得到attention map $V_i$，feature map和attention map进行逐元素相乘后得到SPAB块最终的输出$O_i$，我们使用$W_i^{(j)}\in R^{C_i^{(j)}\times H'\times W'},j=1,2,3$表示第$i$个SPAB块的第$j$层卷积层的参数，SPAB块可以表示如下：
As shown in Figure \ref{fig:pipeline}, SPAN consists of 6 consecutive SPABs and each SPAB block extracts progressively higher-level features sequentially through three convolutional layers with $C'$-channeled $H' \times W'$-sized kernels (In our model, we choose $H' = W' = 3$.). The extracted features $H_i$ are then added with a residual connection from the input of SPAB, forming the pre-attention feature map $U_i$ for that block. The features extracted by the convolutional layers are passed through an activation function $\sigma_a(\cdot)$ that is symmetric about the origin to obtain the attention map $V_i$. The feature map and attention map are element-wise multiplied to produce the final output $O_i=U_i \odot V_i$ of the SPAB block, where $\odot$ denotes element-wise multiplication. We use $W_i^{(j)}\in R^{C' \times H'\times W'}$ to represent the kernel of the $j$-th convolutional layer of the $i$-th SPAB block and $\sigma$ to represent the activation function following the convolutional layer.  Then the SPAB block can be expressed as:
\begin{equation}
    \begin{aligned}
        O_i&=F_{W_i}^{(i)}(O_{i-1})=U_i \odot V_i,\\
        U_i&=O_{i-1}\oplus H_i, \quad V_i=\sigma_a(H_i),\\ 
        H_i&=F_{c,W_i}^{(i)}(O_{i-1}),\\
        &=W_i^{(3)}\otimes\sigma(W_i^{(2)}\otimes\sigma(W_i^{(1)}\otimes O_{i-1})),
    \end{aligned}
    \label{equ:SPAB}
\end{equation}
where $\oplus$ and $\otimes$ represent the element-wise sum between extracted features and residual connections,  and the convolution operation, respectively.  $F_{W_i}^{(i)}$ and $F_{c,W_i}^{(i)}$ are the function representing the $i$-th SPAB and the function representing the $3$ convolution layers of $i$-th SPAB with parameters   $W_i=(W_i^{(1)},W_i^{(2)},W_i^{(3)})$, respectively. $O_0=\sigma(W_0\otimes I_{\text{LR}})$ is a $C'$-channeled $H \times W$ feature map from the $C$-channeled $H \times W$-sized low-resolution input image $I_{LR}$ undergone a convolutional layer with $3 \times 3$ sized kernel $W_0$.  This convolutional layer ensures that each SPAB has the same number of channels as input.  The whole SPAN neural network can be described as
\begin{equation}
    \begin{aligned}
        I_{\text{HR}}&=F(I_{\text{LR}})=\text{PixelShuffle}[W_{f2}\otimes O],\\
        O&=\text{Concat}(O_0,O_1,O_5,W_{f1}\otimes O_6),
    \end{aligned}
    \label{equ:SPAN}
\end{equation}
where $O$ is a $4C'$-channeled $H \times W$-sized feature map with multiple hierarchical features obtaining by concatenating $O_0$ with the outputs of the first, fifth, and the convolved output of the sixth SPAB blocks by $C'$-channeled $3 \times 3$-sized kernel $W_{f1}$.  $O$ is processed through a $3\times 3$ convolutional layer to create an $r^2C$ channel feature map of size $H\times W$. Pixel shuffle is a classic upsampling method~\cite{shi2016real}, which can increase the spatial resolution without adding computational complexity by rearranging the elements in the feature maps. Then, this feature map goes through a pixel shuffle module to generate a high-resolution image of $C$ channels and dimensions $rH\times rW$, where $r$ represents the super-resolution factor. % scaling factor

%$H_0$ 是输入的$H\times W$的低分辨率的图片经过一个 $3 \times 3$卷积层得到的$C$ channel 的$H\times W$ feature map，该feature map使得每个SPAB的输入具有相同的channel数。 In SPAN neural network, the output of 第一个、第五个、第六个 SPAB 块的输出和$H_0$进行连接，形成一个$4C'$ channel的包含了多个层次特征的$H \times W$大小的feature map。$H_c$ 经过一个$3\times 3$的卷积层形成一个$r^2C$ channel的$H\times W$的feature map，然后该feature map经过pixel shuffle模块形成$C$通道的$rH\times rW$的high resolution的图像，这里$r$表示super-resolution 的倍数。
\vspace{-0.5em} 
\subsection{Parameter-Free Attention Mechanism}
\vspace{-0.5em} 
\label{sec:param-free}
%在SPAB中，我们将卷积层提取的高层次特征信息经过关于原点对称的激活函数$\sigma_a$作用直接得到attention map，因为在从计算feature map路径和计算attention map路径的分支处到attention map的路径上除了激活函数外没有包含可训练的参数，所以我们的self-attention是无参数的。在以前的super resolution的工作中，虽然attention map可以提升模型的准确性，by allowing the model to selectively focus on the most relevant parts of the feature，但是对于attention map的计算会引入更多的参数，which 使得模型的计算变慢，为了提高模型的计算性能，我们从\cite{}的无参数的attention map中获得启发，在自注意力机制中通过无参数的激活函数直接获得attention map $V_i$.
In SPAB, we directly obtain the attention map from the higher-level feature information extracted by the convolutional layers through an origin-symmetric activation function $\sigma_a$.  Because in the branch dedicated to computing the attention map that diverges from the computation of the feature map, there are no modules with trainable parameters except for the activation function, our attention mechanism is parameter-free. In previous super-resolution works, while attention maps can improve model accuracy by allowing the model to selectively focus on the most relevant parts of the feature, calculating attention maps introduces additional parameters, which slows down the model's computation~\cite{sun2023safmn, wang2023omni, SwinIR} with extra calculations. To enhance the computational efficiency of our model, we draw inspiration from parameter-free attention mechanisms as proposed in~\cite{du2022parameter,yang2021simam,shi2023parameter,choe2019attention}, and in our attention mechanism, we directly obtain the attention map $V_i$ through a parameter-free activation function $\sigma_a$.  

In the context of the super-resolution task, the reason why computing the attention map in this way is effective lies in the fact that, in super-resolution tasks, when utilizing attention mechanisms, the neural network should focus on where local information such as complex textures, edges, color transitions, and more is particularly rich, where super-resolution is more challenging with issues such as blurring artifacts tending to occur~\cite{ma2020structure}. Interestingly, these edge and texture information required for attention can be directly detected through the convolutional kernels learned during training~\cite{innamorati2020learning} and at the same time, they are also the information that networks need to extract in order to accomplish the super-resolution task. Therefore, we can potentially determine the region for attention directly based on the magnitudes of the convolutional layer's output values and obtain the attention map $V_i=\sigma_a(H_i)$ parameter-free directly from the output of the convolutional layer. This is also reflected in the visualizations of the feature map after using our attention in Figure~\ref{fig:method:1stSPAB} (compared with Figure~\ref{fig:method:1stfeature_no}), where, after training, the attention map $V_i$ directly computed from the output of the convolutional layer $H_i$ tend to make feature map $O_i$  relatively higher in areas with complex textures and boundaries.  
%在super resolution的任务上，我们这样计算attention map有效的原因在于，在super resolution的任务中，若使用attention机制，神经网络关注的relevant parts of the feature应该是局部的纹理、边缘、颜色交界等特征信息，因为在这些边缘、交界、纹理复杂的位置，super resolution较为困难，更容易出现blurring artifact等问题\cite{ma2020structure}，碰巧的是，这些边缘等信息能够通过训练得到的卷积核直接检测得到\cite{innamorati2020learning}，所以，我们可能可以直接通过卷积层的输出的值的大小来判断是否需要attention的依据，从而直接从卷积层的输出无参数的获得attention map，在\autoref{}中的attention map的可视化的图中也反映了，经过训练，卷积层的输出直接计算得到的attention map在纹理复杂、边界等区域的值相对更大

Our parameter-free attention mechanism can be theoretically demonstrated through the following process. It should be noted that because we analyze the role of the residual connection in Section \ref{para:residual}, we remove the residual connection in SPAB for simplicity in this section and add the residual connection for analyzing in Section \ref{para:residual}.  Without attention, the gradient used to update the i-th SPAB during the training of the model can be expressed as 
\begin{equation}
    \begin{aligned}
     \frac{\partial L}{\partial W_i}&=\Pi \frac{\partial F_{W_i}^{(i)}(O_{i-1})}{\partial W_i}\\
     &= \Pi \frac{\partial F_{c,W_i}^{(i)}(O_{i-1})}{\partial W_i},
    \end{aligned}
\end{equation}
where $L$ denotes the loss during training and $\Pi$ represents the product of gradients before $\frac{\partial F_{W_i}^{(i)}(O_{i-1})}{\partial W_i}$ in the gradient chain in back-propagation algorithm.  While after adding self attention mechanism, the gradient is 
%\begin{equation}
%    \begin{aligned}
\begin{align}
        \frac{\partial L}{\partial W_i}&=\Pi \frac{\partial F_{W_i}^{(i)}(O_{i-1})}{\partial W_i}\nonumber\\
        &=\Pi \frac{\partial }{\partial W_i}(F_{c,W_i}^{(i)}(O_{i-1})\odot \sigma_a(F_{c,W_i}^{(i)}(O_{i-1})))\label{equ:gradient_modify}\\
        &=\Pi \frac{\partial F_{c,W_i}^{(i)}(O_{i-1})}{\partial W_i}\odot(H_i\odot \sigma_a'(H_i)+\sigma_a(H_i)).\nonumber 
\end{align}
%    \end{aligned}
%\end{equation}
According to Equation \ref{equ:gradient_modify}, it can be found that for information-richer region, corresponding values in feature $H_i$ and $\sigma_a(H_i)$ have larger absolute value and with $\sigma_a'(H_i)>0$, corresponding absolute value in $H_i\odot \sigma_a'(H_i)+\sigma_a(H_i)$ will be larger to make the information-richer region have more influence on the gradient, so that through the training process, the model will pay more attention to information-rich regions.

%我们的parameter-free的self-attention机制可以通过下面的过程从理论上进行论证。在对模型训练的过程中，更新第i个SPAB使用的梯度可以表示为$\PI \frac{\partial F_{W_i}^{(i)}(O_{i-1})}{\partial W_i}$，其中$PI$表示在使用梯度回传算法计算损失函数关于$W_i$的梯度时，gradient chain中在$\frac{\partial F_{W_i}^{(i)}(O_{i-1})}{\partial W_i}$之前的梯度乘积。

\vspace{-0.5em} 
\subsection{Design Consideration}
\vspace{-0.5em} 
\label{sec:consideration}

The idea of computing attention maps directly without parameters from feature extracted by convolutional layers, led to two design considerations for our neural network: the choice of activation function for computing the attention map and the use of residual connections.

\begin{figure}[!t]
    \centering
\begin{minipage}{0.3\linewidth}\centering    
\includegraphics[width=1\linewidth]{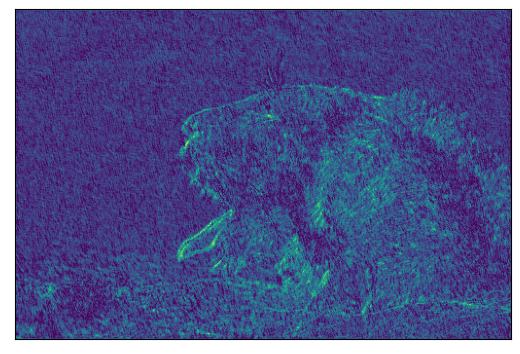}
    \subcaption{$H_1$}\label{fig:method:feature_pre_attention}
\end{minipage}
    \vspace{-0.1in}
\begin{minipage}{0.3\linewidth}\centering    
\includegraphics[width=1\linewidth]{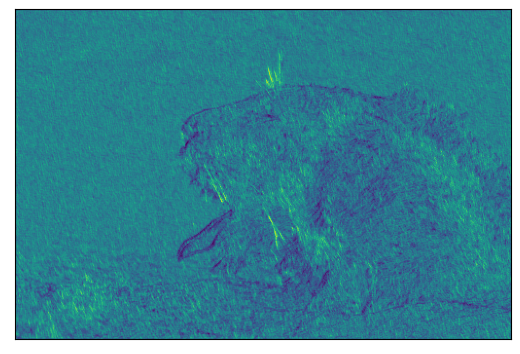}
    \subcaption{$V_1$}\label{fig:method:attention_map}
\end{minipage}
\begin{minipage}{0.3\linewidth}\centering    
\includegraphics[width=1\linewidth]{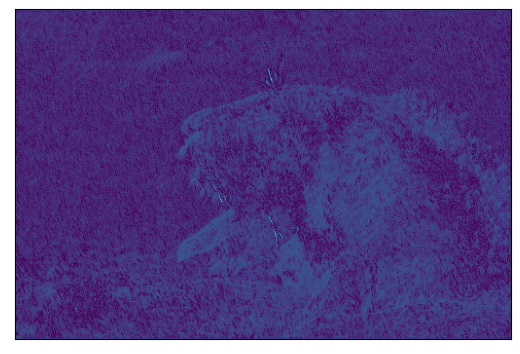}
    \subcaption{$O_1$}
    \label{fig:method:feature_after_attenion}
\end{minipage}
   % \hspace{0.1in}
    \caption{\textbf{Attention Weakened by Non-odd activation function:} From left to right, the sequence are the first SPAB's Pre-attention Feature Map $H_1$, the Attention Map $V_1$, and the Output Feature Map $O_1$, when activation function $\sigma_a$ is sigmoid. In the figures, brighter regions denote larger absolute values.} 
    \label{fig:lmc:odd_function}
    \vspace{-1.8em}
\end{figure}

\noindent\textbf{Symmetric Activation Function} As mentioned in Section \ref{sec:architechture}, we choose the activation function symmetric about the origin to compute the attention map.  There are two main reasons, firstly, because in feature maps extracted through structure-related convolutional layers, like gradient kernels, sign of values always represents directions and the absolute magnitude represents the feature quantity. To generate the attention map $\sigma_a(H_i)$ directly based on feature quantities, it must roughly holds
\begin{equation}
    \begin{aligned}
        &|\sigma_a(x)| = |\sigma_a(|x|)| = |\sigma_a(-x)|.
    \end{aligned}
    \label{equ:first_condition}
\end{equation}
% Second, according to Equation \ref{equ:gradient_modify}, one of the mechanisms of our self-attention method is to enhance the gradient in information-rich region and attenuate the gradient in information-poor region.  To make the effects of enhancement and attenuation obvious, $H_i\odot \sigma_a'(H_i)$ and $\sigma_a(H_i)$ should not cancel each other out, 
Second, per Equation \ref{equ:gradient_modify}, our attention method amplifies gradients in information-rich regions and dampens those in information-poor regions. To ensure these effects, $H_i\odot \sigma_a'(H_i)$ and $\sigma_a(H_i)$ must not cancel each other out, thus it is necessary to ensure:
\begin{equation}
    \begin{aligned}
        x\sigma_a'(x) \cdot \sigma_a(x)>0\\
        \overset{\sigma_a'(x)>0}{\Rightarrow} x \sigma_a(x)>0.
    \end{aligned}
    \label{equ:second_condition}
\end{equation}
The second inequality in Equation \ref{equ:second_condition} is due to the fact that common activation functions are increasing functions. Based on \ref{equ:first_condition} and \ref{equ:second_condition}, it can be deduced that $\sigma_a$ needs to be an odd function symmetric about the origin.  In addition, using non-odd activation functions like Sigmoid, which completely filter out negative values, would result in the loss of information for features with large magnitudes but negative values ($|x|\gg 0,x < 0$), as shown in Figure \ref{fig:lmc:odd_function}.  In Figure \ref{fig:lmc:odd_function}, comparing $H_1$ and $V_1$, it can be observed that in $H_1$, some areas with large absolute feature values become dimmer in $V_1=\sigma_a(H_i)$, representing a decrease in magnitude, due to the effect of the sigmoid function filtering out negative values. The resulting feature map $O_i$, obtained using this type of attention map $V_1$, fails to emphasize the features, resulting in a rather blurred representation compared to initial features.

\begin{figure}[htb]
    \centering
\begin{minipage}{0.3\linewidth}\centering    
\includegraphics[width=1\linewidth]{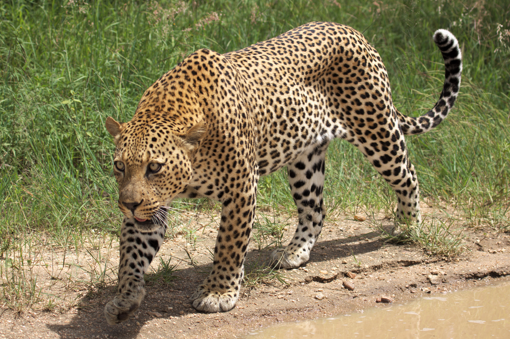}

    \subcaption*{LR Input}
    \label{fig:method:origin1}
\end{minipage}
\\
\begin{minipage}{0.3\linewidth}\centering    
\includegraphics[width=1\linewidth]{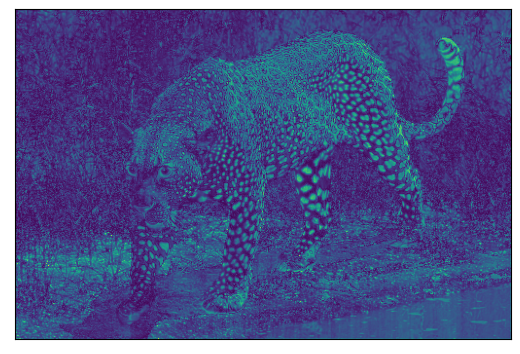}
    \subcaption{1st SPAB}\label{fig:method:1stSPAB}
\end{minipage}
    %\hspace{0.1in}
\begin{minipage}{0.3\linewidth}\centering    
\includegraphics[width=1\linewidth]{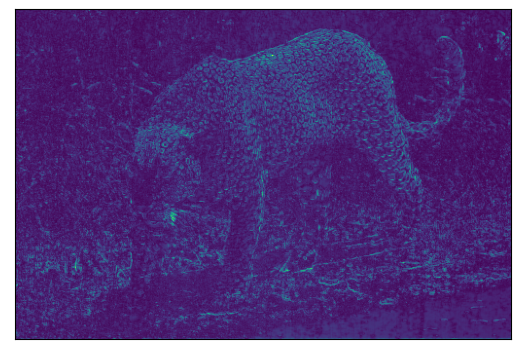}
    \subcaption{1st SPAB\_nores}
    \label{fig:method:1stfeature_nores}
\end{minipage}
\begin{minipage}{0.3\linewidth}\centering    
\includegraphics[width=\linewidth]{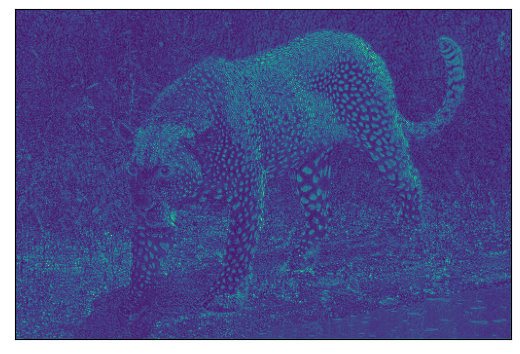}
    \subcaption{1st SPAB\_empty}
    \label{fig:method:1stfeature_no}
\end{minipage}
\begin{minipage}{0.3\linewidth}\centering    
\includegraphics[width=1\linewidth]{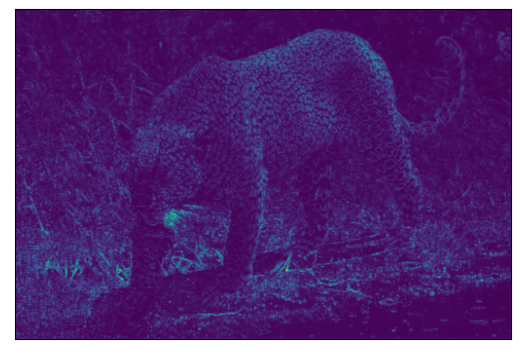}
    \subcaption{6th SPAB}\label{fig:method:6thSPAB}
\end{minipage}
\begin{minipage}{0.3\linewidth}\centering    
\includegraphics[width=\linewidth]{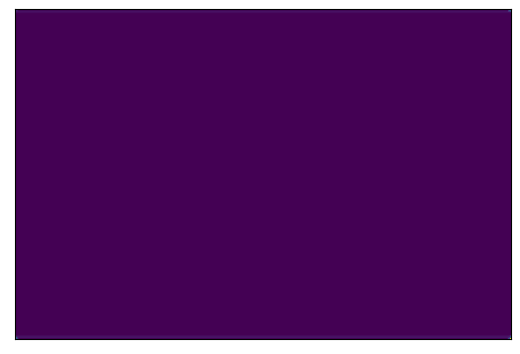}
    \subcaption{6th SPAB\_nores}
    \label{fig:method:6thfeature_nores}
\end{minipage}
\begin{minipage}{0.3\linewidth}\centering    
\includegraphics[width=\linewidth]{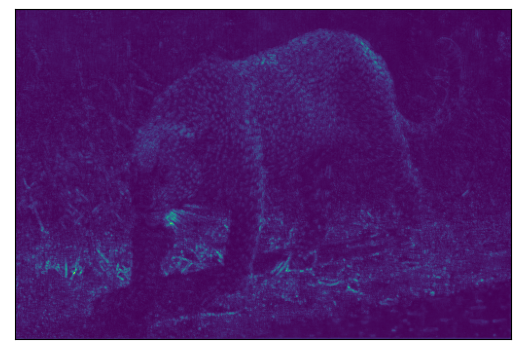}
    \subcaption{6th SPAB\_empty}
\label{fig:method:6thfeature_no}
\end{minipage}
    \caption{\textbf{Residual Connections Improve Attention:} a comparison is made between the output feature maps of the first and sixth SPABs for complete SPAB \textcolor{red}{(a) (d)}, SPAB with no residual connections \textcolor{red}{(b) (e)}, and SPAB without attention and residual connections \textcolor{red}{(c) (f)}. It is observed that the SPAB with attention but lacking residual connections tends to lose a substantial amount of information, which issue is alleviated by incorporating residual connections.}
    \label{fig:lmc:model_archs_landscape}
    \vspace{-1.8em}
\end{figure}

\begin{table*}[htbp]
% \vspace{1em}
    % \renewcommand{\arraystretch}{1.2}
    \centering
        
    \vspace{-1.em}
    % \scalebox{0.65}{
    \resizebox{1.0\linewidth}{!}{
        \begin{tabular}{c c c c c c c c c}
            \toprule
            \multirow{2}{*}{Scale} & \multirow{2}{*}{Model} & Params & Runtime & Set5 & Set14 & BSD100 & Urban100 & Manga109 \\
            ~ & ~ & (K) & (ms) &PSNR$\uparrow$ / SSIM$\uparrow$ & PSNR$\uparrow$ / SSIM$\uparrow$ & PSNR$\uparrow$ / SSIM$\uparrow$ & PSNR$\uparrow$ / SSIM$\uparrow$ & PSNR$\uparrow$ / SSIM$\uparrow$ \\
     
            \midrule
            \multirow{15}{*}{$\times$ 2} & SRCNN\cite{dong2014learning} & 24 & 6.92
            & 36.66 / 0.9542
            & 32.42 / 0.9063
            & 31.36 / 0.8879
            & 29.50 / 0.8946 
            & 35.74/0.9661 \\
            
            ~ & FSRCNN\cite{FSRCNN} & 12 & 9.02
            & 36.98 / 0.9556
            & 32.62 / 0.9087
            & 31.50 / 0.8904
            & 29.85 / 0.9009 
            & 36.67/0.9694 \\
            
            ~ & VDSR\cite{VDSR} & 666 & 35.37
            & 37.53 / 0.9587
            & 33.05 / 0.9127
            & 31.90 / 0.8960
            & 30.77 / 0.9141 
            & 37.22/0.9729 \\
            
            ~ & DRCN\cite{DRCN} & 1774 & 716.45 
            & 37.63 / 0.9588 
            & 33.04 / 0.9118 
            & 31.85 / 0.8942 
            & 30.75 / 0.9133
            & 28.93 / 0.8854 \\
            
            ~ & LapSRN\cite{LapSRN} & 251 & 53.98 
            & 37.52 / 0.9591 
            & 32.99 / 0.9124 
            & 31.80 / 0.8952 
            & 30.41 / 0.9103 
            & 37.27/0.9740 \\
            
            ~ & CARN\cite{CARN} & 1592 & 159.10 
            & 37.76 / 0.9590 
            & 33.52 / 0.9166 
            & 32.09 / 0.8978 
            & 31.92 / 0.9256 
            & 38.36/0.9765 \\
            
            ~ & IMDN\cite{IMDN} & 694 & 77.34 
            & 38.00 / 0.9605 
            & 33.63 / 0.9177 
            & 32.19 / 0.8996 
            &  32.17 / 0.9283
            & \textcolor{blue}{38.88}/0.9774 \\
            
            ~ & RFDN\cite{RFDN} & 534 & 74.51 
            & 38.05 / 0.9606 
            & 33.68 / 0.9184 
            & 32.16 / 0.8994 
            & 32.12 / 0.9278 
            & \textcolor{blue}{38.88}/0.9773 \\
            
            % ~ & RFDN-L\cite{RFDN} & 626
            % & 38.08 / 0.9606 
            % & 33.67 / 0.9190 
            % & 32.18 / 0.8996 
            % & 32.24 / 0.9290 \\
            % % & 30.61 / 0.9096 \\
            
            ~ & MAFFSRN\cite{MAFFSRN} & 402 & 152.91 
            & 37.97 / 0.9603 
            & 33.49 / 0.9170 
            & 32.14 / 0.8994 
            & 31.96 / 0.9268 
            &  /  \\
            
            ~ & ECBSR\cite{ECBSR} & 596 & 39.96
            & 37.90 / \textcolor{red}{0.9615}
            & 33.34 / 0.9178
            & 32.10 / \textcolor{red}{0.9018}
            & 31.71 / 0.9250 
            &  /  \\

            ~ & RLFN-S\cite{kong2022residual} & 454 & 56.09 
            & 38.05 / 0.9607
            & 33.68 / 0.9172
            & 32.19 / 0.8997
            & 32.17 / 0.9286 
            &  /  \\
            
            ~ & RLFN\cite{kong2022residual} & 527 & 60.39  
            & \textcolor{blue}{38.07} / 0.9607
            & \textcolor{blue}{33.72} / \textcolor{red}{0.9187}
            & \textcolor{red}{32.22} / 0.9000
            & \textcolor{red}{32.33} / \textcolor{red}{0.9299} 
            &  /  \\
            
            ~ & ShuffleMixer\cite{sun2022shufflemixer} & 394 & 218.36 
            & 38.01 / 0.9606 
            & 33.63 / 0.9180 
            & 32.17 / 0.8995
            & 31.89 / 0.9257 
            & 38.83/0.9774 \\

            ~ & SAFMN\cite{sun2023safmn} & 228 & 118.07 
            & 38.00 / 0.9605 
            & 33.54 / 0.9177 
            & 32.16 / 0.8995
            & 31.84 / 0.9256 
            & 38.71/0.9771 \\
            
            ~ & \textbf{SPAN-S (ours)} & 411 & \textbf{45.08}
            & 38.06 / \textcolor{blue}{0.9608}
            & \textcolor{red}{33.73} / \textcolor{red}{0.9187}
            & \textcolor{blue}{32.21} / 0.9001
            & 32.20 / 0.9288 
            & 38.85 / \textcolor{blue}{0.9776}  \\
            
            ~ & \textbf{SPAN (ours)} & 481 & \textbf{50.39}  
            & \textcolor{red}{38.08} / \textcolor{blue}{0.9608}
            & 33.71 / \textcolor{blue}{0.9183}
            & \textcolor{red}{32.22} / \textcolor{blue}{0.9002}
            & \textcolor{blue}{32.24} / \textcolor{blue}{0.9294} 
            & \textcolor{red}{38.94} / \textcolor{red}{0.9777}  \\
        
            % \midrule
            
            \midrule
            \multirow{16}{*}{$\times$ 4} & SRCNN\cite{dong2014learning} & 57 & 1.90
            & 30.48 / 0.8628
            & 27.49 / 0.7503
            & 26.90 / 0.7101
            & 24.52 / 0.7221 
            & 27.58 / 0.8555 \\
            
            ~ & FSRCNN\cite{FSRCNN} & 13 & 2.22
            & 30.72 / 0.8660
            & 27.61 / 0.7550
            & 26.98 / 0.7150
            & 24.62 / 0.7280 
            & 27.90 / 0.8610 \\
            
            ~ & VDSR\cite{VDSR} & 666 & 8.95
            & 31.35 / 0.8838 
            & 28.01 / 0.7674
            & 27.29 / 0.7251
            & 25.18 / 0.7524 
            & 28.83 / 0.8870 \\
            
            ~ & DRCN\cite{DRCN} & 1774 & 176.59 
            & 31.53 / 0.8854
            & 28.02 / 0.7670 
            & 27.23 / 0.7233
            & 25.14 / 0.7510 
            & 28.93 / 0.8854 \\
            
            ~ & LapSRN\cite{LapSRN} & 502 & 66.81 
            & 31.54 / 0.8852 
            & 28.09 / 0.7700
            & 27.32 / 0.7275 
            & 25.21 / 0.7562 
            & 29.09 / 0.8900 \\
            
            ~ & CARN\cite{CARN} & 1592 & 39.96 
            & 32.13 / 0.8937 
            & 28.60 / 0.7806 
            & 27.58 / 0.7349 
            & 26.07 / 0.7837 
            & 30.47 / 0.9084 \\
            
            ~ & IMDN\cite{IMDN} & 715 & 20.56 
            & 32.21 / 0.8948 
            & 28.58 / 0.7811 
            & 27.56 / 0.7353 
            & 26.04 / 0.7838 
            & 30.45 / 0.9075 \\
            
            ~ & RFDN\cite{RFDN} & 550 & 20.40 
            & 32.24 / 0.8952
            & 28.61 / 0.7819
            & 27.57 / 0.7360
            & 26.11 / 0.7858 
            & 30.58 / 0.9089 \\
            
            % ~ & RFDN-L\cite{RFDN} & 643K
            % & 32.28 / 0.8957
            % & 28.61 / 0.7818 
            % & 27.58 / 0.7363
            % & 26.20 / 0.7883 \\
            % % & 30.61 / 0.9096 \\
            
            ~ & MAFFSRN\cite{MAFFSRN} & 441 & 39.69 
            & 32.18 / 0.8948 
            & 28.58 / 0.7812 
            & 27.57 / 0.7361
            & 26.04 / 0.7848 
            &  / \\
            
            ~ & ECBSR\cite{ECBSR} & 603 & 10.21
            & 31.92 / 0.8946 
            & 28.34 / 0.7817 
            & 27.48 / \textcolor{red}{0.7393}
            & 25.81  /0.7773 
            &  / \\

            ~ & FDIWN-M\cite{FDIWN} & 454 & - 
            & 32.17 / 0.8941 
            & 28.55 / 0.7806 
            & 27.58 / 0.7364
            & 26.02 / 0.7844 
            &  / \\
            
            % ~ & FDIWN\cite{FDIWN} & 664
            % & 32.23 / 0.8955 
            % & 28.66 / 0.7829 
            % & 27.62 / 0.7380 
            % & 26.28 / 0.7919 \\
            % ~ & RepSR\cite{wang2022repsr} & 630 & - 
            % & 31.94 / \textcolor{red}{0.8961}
            % & 28.44 / 0.7853
            % & 27.48 / 0.7825
            % & 25.88 / 0.7825 
            % &  / \\
            
            ~ & RLFN-S\cite{kong2022residual} & 470 & 15.16 
            & \textcolor{blue}{32.23} / \textcolor{red}{0.8961}
            & 28.61 / 0.7818
            & 27.58 / 0.7359
            & 26.15 / 0.7866 
            &  / \\
            
            ~ & RLFN\cite{kong2022residual} & 543 & 16.41 
            & \textcolor{red}{32.24} / 0.8952
            & 28.62 / 0.7813
            & 27.60 / 0.7364
            & \textcolor{blue}{26.17} / \textcolor{blue}{0.7877} 
            &  / \\

            ~ & ShuffleMixer\cite{sun2022shufflemixer} & 411 & 144.22 
            & 32.21 / \textcolor{blue}{0.8953}
            & 28.66 / 0.7827 
            & \textcolor{blue}{27.61} / 0.7366
            & 26.08 / 0.7835 
            & \textcolor{blue}{30.65} / 0.9093 \\

            ~ & SAFMN\cite{sun2023safmn} & 240 & 72.06 
            & 32.18 / 0.8948 
            & 28.60 / 0.7813 
            & 27.58 / 0.7359
            & 25.97 / 0.7809 
            & 30.43/0.9063 \\

            ~ & \textbf{SPAN-S (ours)} & 426 & \textbf{12.22}
            & 32.20 / 0.8950
            & \textcolor{blue}{28.64} / \textcolor{blue}{0.7828}
            & 27.60 / 0.7368
            & 26.13 / 0.7865 
            & 30.60 / \textcolor{blue}{0.9095} \\
            
            ~ & \textbf{SPAN (ours)} & 498 & \textbf{13.67}
            &  32.20 / \textcolor{blue}{0.8953}
            & \textcolor{red}{28.66} / \textcolor{red}{0.7834}
            & \textcolor{red}{27.62} / \textcolor{blue}{0.7374}
            & \textcolor{red}{26.18} / \textcolor{red}{0.7879} 
            & \textcolor{red}{30.66} / \textcolor{red}{0.9103} \\
            
            \bottomrule
            \end{tabular}
    }
    % }
\caption{Quantitative results of the state-of-the-art ESR models on five benchmark datasets. The approach to evaluating inference time remains consistent with RLFN~\cite{kong2022residual}. The best and second-best results are marked in \textcolor{red}{red} and \textcolor{blue}{blue} colors, and the \textbf{bold} numbers represent that the inference speed of our model is the fastest when the performance and number of parameters are similar.}
\label{tab:quantitative_sota}

\vspace{-1.5em}
\end{table*}
\noindent\textbf{Residual Connection}\label{para:residual} On the other hand, the attention mechanism enhances higher-level features extracted at each layer, which can lead to a significant loss of information in regions where higher-level features are less prominent. Compared to the results without using attention (Figure \ref{fig:method:6thfeature_no}), using attention can cause issue of the excessive loss of information with the later SPAB block (Figure \ref{fig:method:6thfeature_nores}), ultimately leading to a decrease in the accuracy of super-resolution results (Table \ref{tab:ablation_three}). We employ residual connections to address this problem. Through residual connections, we use lower-level features from the input of the SPAB layer to compensate for the excessive loss of information during the feature map generation process, and in the final SPAB (Equation \ref{equ:SPAB}), $V_i\odot H_i$ is replaced by $V_i\odot (H_i\oplus O_{i-1})$. This will result in the gradient $\frac{\partial L}{\partial W_i}$ for updating $W_i$ during the training process in Equation \ref{equ:gradient_modify} become
    \begin{align}
        &\frac{\partial L}{\partial W_i}=\Pi \frac{\partial F_{W_i}^{(i)}(O_{i-1})}{\partial W_i}\nonumber\\
        &\!=\!\Pi \!\frac{\partial }{\partial W_i}\!((F_{c,W_i}^{(i)}(O_{i-1})\!+\!O_{i-1})\!\odot\! \sigma_a(F_{c,W_i}^{(i)}(O_{i-1})))\\
        &\!=\!\Pi \frac{\partial F_{c,W_i}^{(i)}(O_{i-1})}{\partial W_i}\!\odot\!((H_i\!+\!O_{i-1})\!\odot\! \sigma_a'(H_i)\!+\!\sigma_a(H_i))\nonumber.
    \end{align}
    In this way,  because of the existing of $O_{i-1}$ in $\frac{\partial L}{\partial W_i}$, for the trained model, the regions focused on by the attention mechanism are determined not only by the information of the current level but also by the lower-level information output by the previous SPAB. This alleviates the significant loss of information caused by overly focusing on higher-level features.  
    
    By comparing the feature maps of the first and sixth layers of the block with no attention (Figure \ref{fig:method:1stfeature_no} and \ref{fig:method:6thfeature_no}), the block with attention (Figure \ref{fig:method:1stfeature_nores} and \ref{fig:method:6thfeature_nores}), and the block with improved attention through residual connections (Figure \ref{fig:method:1stSPAB} and \ref{fig:method:6thSPAB}), we observe that the SPAB block with residual connections can further highlight features compared to the block with no attention. At the same time, it does not suffer from the loss of excessive information seen in the block with attention but without residual connections, ensuring that lower-level information is preserved. Detailed numerical results are shown in Table \ref{tab:ablation_three}.
    \vspace{-0.7em}

\section{Experiments}
\vspace{-0.5em}
\subsection{Experimental Setup} %这里数据集描述部分缺少对benchmark使用的描述，以及对测试指标的方法描述
\vspace{-0.5em}
\noindent\textbf{Datasets and Metrics} In accordance with established techniques~\cite{LAPAR, SwinIR}, our models are trained on the DF2K dataset, which is a combination of DIV2K~\cite{DIV2K} and Flickr2K~\cite{EDSR} datasets, comprising a total of 3450 (800 + 2650) high-quality images.
We adopt standard protocols to generate LR images by bicubic downscaling of reference HR images. 
We evaluate our models on 4 different benchmark dataset: Set5~\cite{Set5}, Set14~\cite{Set14}, B100 ~\cite{BSD100}, Urban100~\cite{Urban100} and Manga109~\cite{matsui2017sketch}, PSNR and SSIM are used as performance on the Y channel of YCbCr space for SR task. \\
\noindent\textbf{Implementation Details}
%The training procedure entails the deployment of 6 SPAB modules arranged in a sequence, each consisting of 48 feature maps.
Six SPAB modules with 48-channeled feature maps are depolyed for training procedure.  We have employed the re-parameterization method (REP) \cite{ding2021repvgg} to improve the efficiency during the inference stage. During each training batch, 64 HR RGB patches are cropped, measuring $256 \times 256$, and subjected to random flipping and rotation. The learning rate is initialized at $5\times 10^{-4}$ and undergoes a halving process every $2\times10^5$ iterations. The network undergoes training in for a total of $10^6$ iterations on four NVIDIA A100 GPUs, with each run taking approximately 12 hours, and the L1 loss function being minimized through the utilization of the Adam optimizer\cite{kingma2014adam}. We perform the aforementioned training settings twice after loading the trained weights to obtain the optimal results.

\vspace{-1.em} 
\subsection{Quantitative Results}
\vspace{-1.em} 
%我们将我们的span和span-s在不同benchmark上2倍和4倍上采样的具体测试结果与现在最先进的高效超分模型进行了对比，结果细节请参见表2，SPAN和SPAN-S的结果在多个benchmark上与其他模型对比PSNR和SSIM指标上性能表现最好，尤其是在inference time方面，SPAN和SPAN-S在参数量方面少50K的前提下来对比RLFN和RLFN-S，在推理速度和模型性能指标PSNR和SSIM上具有明显优势。如图1所示，在可视化模型性能，推理时间和模型大小的关系后，我们很容易发现在近似推理速度下，SPAN的PSNR均远超其他模型的，而在近似模型参数量时，模型不仅性能更好，速度也更快，因此RLFN和RLFN-S在性能，参数和推理速度上实现了当前最好的均衡。

In this study, we conduct 2x and 4x upscaling on the SPAN and SPAN-S models across various benchmark tests, and compare their detailed test results with the current state-of-the-art efficient super-resolution models~\cite{dong2014learning, FSRCNN, VDSR, DRCN, LapSRN, CARN, IMDN, RFDN, MAFFSRN, ECBSR, FDIWN, wang2022repsr, kong2022residual, sun2022shufflemixer, sun2023safmn}. 
For detailed results, please refer to Table~\ref{tab:quantitative_sota}. Across multiple benchmarks, SPAN and SPAN-S exhibit superior performance in terms of PSNR and SSIM compared to other models, especially notable in inference time. With 50K fewer parameters than RLFN and RLFN-S, SPAN and SPAN-S demonstrate significant advantages in inference speed and in the performance metrics of PSNR and SSIM. As illustrated in Figure~\ref{fig:comparison}, through visualizing the relationship between image quality, inference time, and model size, we observe that SPAN achieves significantly higher PSNR than other models at comparable inference speeds; and with similar model parameter counts, it not only performs better but also operates faster. Hence, RLFN and RLFN-S have achieved the best current balance in terms of quality, parameter count, and inference speed.

\vspace{-0.5em} 
\subsection{Activation Function}
\vspace{-0.5em} 
%在之前的method章节，我们提到使用关于原点对称的激活函数对无参数attention机制的作用和原理，我们尝试并也提出了很多不同的关于原点对称的激活函数并比较他们作用于attention机制时的超分性能，如图4所示，我们设置实验对比了不同激活函数的作用后，为了保证模型的速度，最终使用了sigmoid-0.5作为我们对SPAN模型中attention机制的激活函数设计，而提出的一些可学习的激活函数作为SPAN模型的延伸和激活函数的探讨，可以发现可学习的激活函数的性能可能会更高，但速度也会相对下降。
% 画图和表

\begin{table}[!t]
% \vspace{1em}
% \renewcommand{\arraystretch}{1.2}
    \centering
    % \vspace{-0.5em}
    \scalebox{0.61}{
        \begin{tabular}{c c c c c }
            \toprule
            \multirow{1}{*}{$\sigma_a(x)$} & Learnable & Set14 & BSD100 & Urban100 \\
            ~ & ~ & PSNR/SSIM & PSNR/SSIM & PSNR/SSIM \\
            
            \midrule

            $\text{Sigmoid}(x)$ &  -
            &  28.62/0.7826
            &  27.59/0.7366
            &  26.08/0.7854  \\
            
            $\text{Sigmoid}(x)-0.5$  &  -
            &  28.63/0.7825
            &  27.60/0.7368
            &  26.10/0.7856  \\

            % $\text{Tanh}(x)$\cite{tanh} &  -
            % &  28.60/0.7820
            % &  27.59/0.7363
            % &  26.05/0.7841  \\

            $\text{Sigmoid}(ax)-0.5$ &  $\checkmark$            
            &  28.62/0.7825
            &  27.60/0.7368
            &  26.12/0.7861 \\

            $b\times(\text{Sigmoid}(ax)-0.5)$ & $\checkmark$
            &  28.62/0.7826
            &  27.61/0.7367
            &  26.11/0.7860 \\
            
            \bottomrule
        \end{tabular}
    }
    \caption{Performance comparison of using different activation functions $\sigma_a(x)$ evaluated on three benchmark datasets.}
    \label{tab:ablation_activation}
    \vspace{-1.8em}
\end{table}

In the Section~\ref{sec:param-free}, we discuss the use of origin-symmetric activation functions $\sigma_a(x)$ in the context of a parameter-free attention mechanism, its role and principles. We experiment with and proposed several different origin-symmetric activation functions, comparing their impact on the ESR performance with the attention mechanism. As shown in Table~\ref{tab:ablation_activation}, we conduct experiments to compare the effects of different activation functions. To ensure model speed, we ultimately choose $\sigma_a(x)=\text{Sigmoid}(x)-0.5$ as the activation function, which is simple but effective for the attention mechanism in our SPAN model. The proposed learnable activation functions, serving as an extension to the SPAN model and a discussion point on activation functions, demonstrate potentially higher performance. However, it is observed that the speed tends to decrease with the use of learnable activation functions.

\vspace{-0.5em} 
\subsection{Ablation Study}
\vspace{-0.5em} 
%这一节我们进行消融实验，将
In the ablation experiment, we uniformly conduct x4 scale factor experiments on the models with 48 channels. \\
\noindent\textbf{Residual Connection}
\label{sec:ablation-res}
As discussed in Section~\ref{sec:model-arch}, we have improved the performance of our model under the non-parametric attention mechanism by incorporating residual connections within each SPAB module to mitigate excessive information loss. To demonstrate the effectiveness of these intra-module residual connections, we conduct experiments by removing them from every SPAB module in our model and comparing the results with our baseline model. The model without residual connections is denoted as SPAN\_nores. All experimental settings for this model are kept consistent with our SPAN model.

We thoroughly train both models under identical settings and evaluate their performance on four benchmark datasets: Set5, Set14, BSD100, and Urban100. Table~\ref{tab:ablation_three} presents the results, highlighting the impact of residual connections on model performance. Notably, these connections enhance image quality while maintaining high inference speed.

% Table~\ref{tab:ablation_three} presents the results, showcasing the impact of residual connections within the modules on the overall model performance. Notably, the inclusion of residual connections leads to improved performance while maintaining a relatively high inference speed.

\begin{table}[!t]
% \vspace{1em}
% \renewcommand{\arraystretch}{1.2}
    \centering

    % \vspace{-0.5em}
    \scalebox{0.61}{
        \begin{tabular}{c c c c c c }
            \toprule
            \multirow{2}{*}{Model} & {Runtime} & Set5 & Set14 & BSD100 & Urban100 \\
            ~ & (ms) & PSNR / SSIM & PSNR / SSIM & PSNR / SSIM & PSNR / SSIM\\
            
            \midrule
            
            SPAN\_nores  &  11.65
            &  31.25/0.8808
            &  28.05/0.7687
            &  27.21/0.7236
            &  25.09/0.7496  \\
            
            SPAN\_noatt &  10.85
            &  32.07/0.8943
            &  28.56/0.7813
            &  27.57/0.7357
            &  25.97/0.7817  \\

            SPAN\_empty &  10.22
            &  32.06/0.8939 
            &  28.56/0.7811 
            &  27.56/0.7355 
            &  26.02/0.7830  \\
            
            SPAN (ours) & 12.22
            &  32.18/0.8950
            &  28.63/0.7825
            &  27.60/0.7368
            &  26.10/0.7856  \\
                        
            \bottomrule
        \end{tabular}
    }
\caption{Performance comparison of SPAN\_nores, SPAN\_noatt, SPAN\_empty, and our baseline SPAN. Runtime is the average time of 10 runs on DF2K validation set.}
\label{tab:ablation_three}
\vspace{-1.5em}
\end{table}

\newlength{\imgwidth}
\setlength{\imgwidth}{0.12\textwidth} % 设置图像宽度
\begin{figure*}[!t]
\begin{center}
\scalebox{1}{
\renewcommand{\arraystretch}{0.8}
\begin{tabular}[b]{c@{ } c@{ }  c@{ } c@{ } c@{ } c@{ }	}
\hspace{-2mm}
\multirow{4}{*}{%
    \includegraphics[height=29.8mm,width=44.5mm,valign=t]{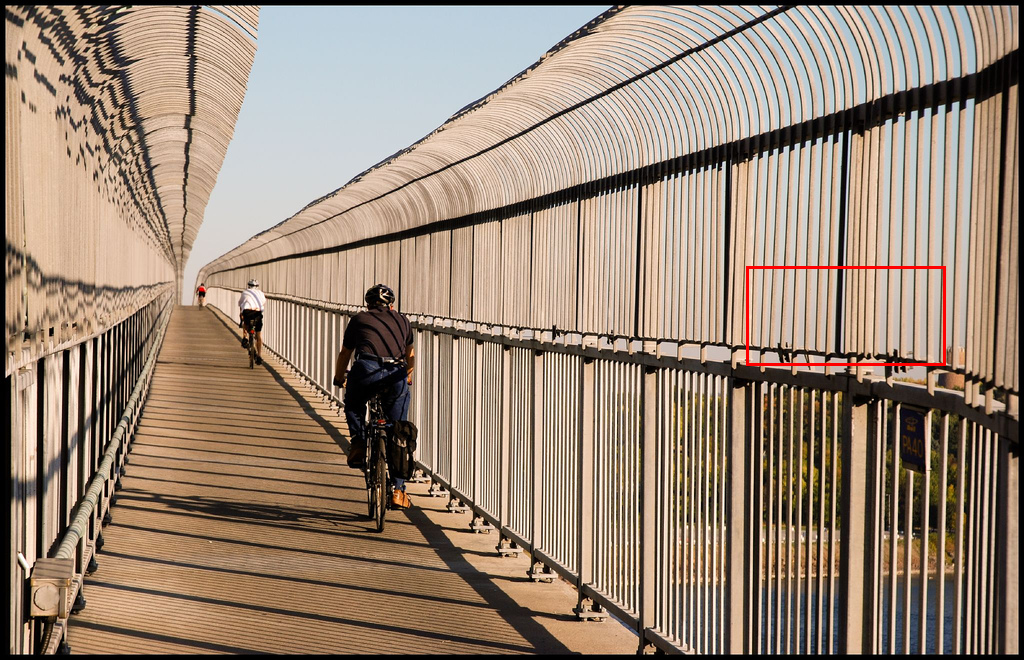}
} &  
    \includegraphics[width=0.135\textwidth,valign=t]{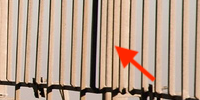}&
    \includegraphics[width=0.135\textwidth,valign=t]{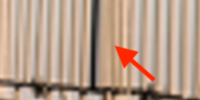}&
    \includegraphics[width=0.135\textwidth,valign=t]{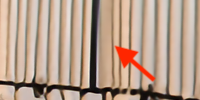}&
    \includegraphics[width=0.135\textwidth,valign=t]{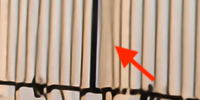}&
    \includegraphics[width=0.135\textwidth,valign=t]{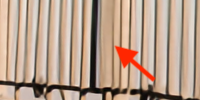}
\\
    &  \small~PSNR &\small~18.06 dB  & \small~21.15 dB & \small~21.53 dB & \small~20.48 dB   \\
    
    & \small~Reference & \small~Bicubic    & \small~CARN-M~\cite{CARN}& \small~CARN~\cite{CARN}& \small~IMDN~\cite{IMDN} \\

    &

    \includegraphics[width=0.135\textwidth,valign=t,valign=t]{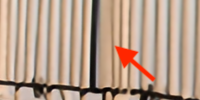}&  
     \includegraphics[width=0.135\textwidth,valign=t]{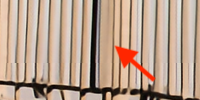}&
     \includegraphics[width=0.135\textwidth,valign=t]{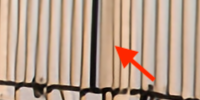}&
    \includegraphics[width=0.135\textwidth,valign=t]{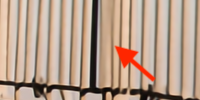}&
     \includegraphics[width=0.135\textwidth,valign=t]{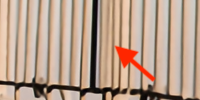}\\

     \multirow{2}{*}{Urban100: \texttt{img\_024.png}} & \small~19.75 dB & \small~21.20 dB & \small 21.26 dB  & \small 21.54 dB & \small~\textbf{22.26 dB}\\
     & \small~ShuffleMixer~\cite{sun2022shufflemixer} & \small~FDIWN~\cite{FDIWN}   & \small~SAFMN~\cite{sun2023safmn} & \small~EDSR~\cite{EDSR} & \small~\textbf{SPAN (Ours)}

\\
\hspace{-2mm}
\multirow{4}{*}{%
    \includegraphics[height=29.8mm,width=44.5mm,valign=t]{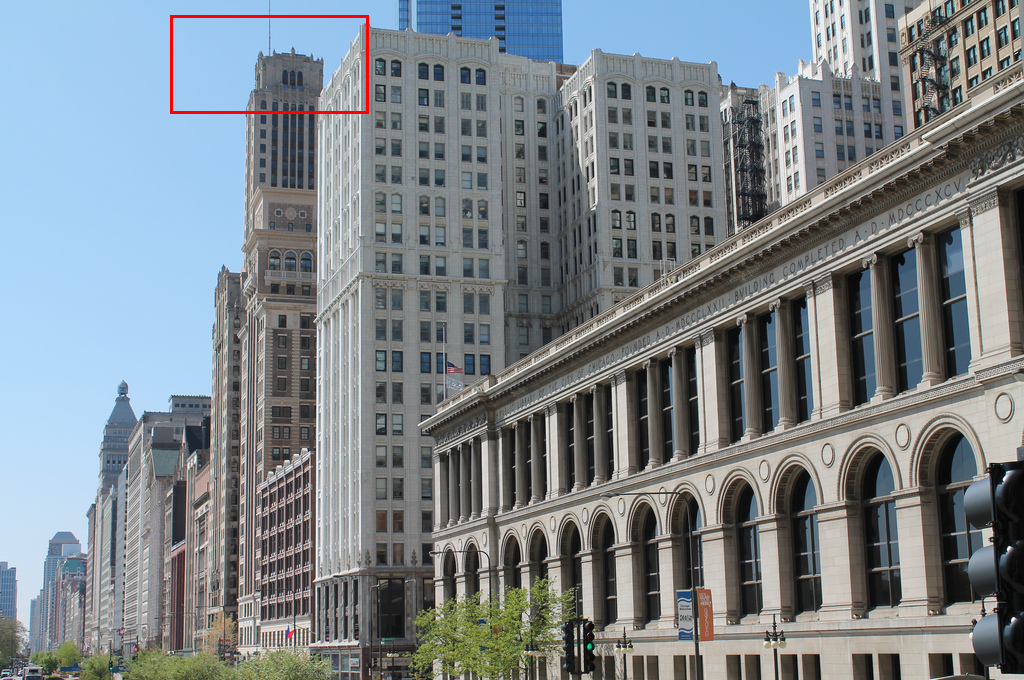}
} &  

    \includegraphics[width=0.135\textwidth,valign=t]{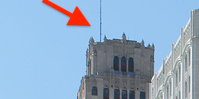}&
    \includegraphics[width=0.135\textwidth,valign=t]{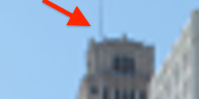}&
    \includegraphics[width=0.135\textwidth,valign=t]{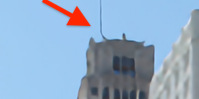}&
    \includegraphics[width=0.135\textwidth,valign=t]{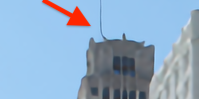}&
    \includegraphics[width=0.135\textwidth,valign=t]{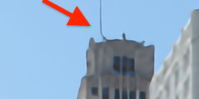}
\\
    &  \small~PSNR &\small~28.14 dB  & \small~29.24 dB & \small~29.33 dB & \small~29.46 dB   \\
    
    & \small~Reference & \small~Bicubic    & \small~CARN-M~\cite{CARN}& \small~CARN~\cite{CARN} & \small~IMDN~\cite{IMDN} \\

    &

    \includegraphics[width=0.135\textwidth,valign=t,valign=t]{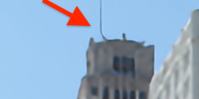}&  
     \includegraphics[width=0.135\textwidth,valign=t]{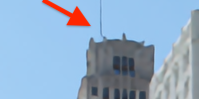}&
     \includegraphics[width=0.135\textwidth,valign=t]{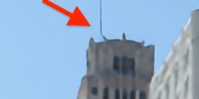}&
    \includegraphics[width=0.135\textwidth,valign=t]{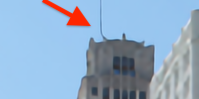}&
     \includegraphics[width=0.135\textwidth,valign=t]{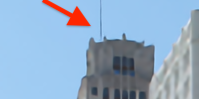}\\

     \multirow{2}{*}{Urban100: \texttt{img\_097.png}} & \small~29.42 dB & \small~29.54 dB & \small 29.33 dB  & \small 29.59 dB & \small~\textbf{29.71 dB}\\
     & \small~ShuffleMixer~\cite{sun2022shufflemixer} & \small~FDIWN~\cite{FDIWN}   & \small~SAFMN~\cite{sun2023safmn}& \small~EDSR~\cite{EDSR} & \small~\textbf{SPAN (Ours)}
\end{tabular}
}
\end{center}
% \vspace{-6mm}
\caption{Visual comparison for $\times4$ SR methods. The patches for comparison are marked with \red{red} boxes.
% in the original images. Compared to other approaches, our Omni-SR generates better visual result with less artifacts.
(Best viewed by zooming.)
}
\label{fig:vis}\vspace{-1em}
\end{figure*}

\noindent\textbf{Attention Mechanism}
\label{sec:ablation:att}
To assess the efficacy of our attention mechanism's foundation, which utilizes a parameter-free attention process for information feedback, we conduct an experiment by removing the activation function and the element-wise multipication operation from the attention component within the SPAB module. This effectively eliminates the parameter-free attention mechanism, enabling the module to output a direct residual connection with the original input. Maintaining the same experimental setup, we designate this modified model as SPAN\_noatt and compare it across various dimensions with the original SPAN model. Table~\ref{tab:ablation_three} demonstrates the enhanced performance of SPAN over SPAN\_noatt, validating the effectiveness of our parameter-free attention mechanism in augmenting network capabilities and accentuating high-frequency features in feature maps.

% As demonstrated in Table~\ref{tab:ablation_three}, SPAN shows improved performance over SPAN\_noatt, confirming that our design of a parameter-free attention mechanism can effectively enhance network capabilities. This improvement notably underscores the mechanism's ability to emphasize high-frequency features within feature maps.

% \begin{table}[!htb]
% % \vspace{1em}
% % \renewcommand{\arraystretch}{1.2}
%     \centering
%     \scalebox{0.55}{
%         \begin{tabular}{c c c c c c c }
%             \toprule
%             \multirow{2}{*}{Model} & Params & {Runtime} & Set5 & Set14 & BSD100 & Urban100 \\
%             ~ & (K) & (ms) & PSNR / SSIM & PSNR / SSIM & PSNR / SSIM & PSNR / SSIM\\
            
%             \midrule
            
%             SPAN\_noatt &  &  
%             &  / 
%             &  / 
%             &  / 
%             &  /  \\
            
%             SPAN (ours) &  & 
%             &  / 
%             &  / 
%             &  / 
%             &  /  \\
            
%             \bottomrule
%         \end{tabular}
%     }
%     \caption{Comparision of SPAN\_noatt and our baseline SPAN with  Runtime is the average of 10 runs on DIV2K validation set.}
%     \label{tab:ablation_att}
% \vspace{-1.5em}
% \end{table}

\noindent\textbf{Combination Module}
\label{sec:ablation-combine}
We also perform an extensive validation of the proposed module design. The SPAB module, which integrates residual connections with a parameter-free attention mechanism, achieves efficient direct enhancement. To evaluate this, we remove these two key components in our experiments while maintaining other experimental settings constant. Results from Table~\ref{tab:ablation_three} demonstrate that our proposed attention module, incorporating residual connections, significantly improves image quality in super-resolution tasks without compromising processing speed. These findings robustly substantiate the effectiveness and practicality of our module design in enhancing super-resolution processing outcomes.

\noindent\textbf{Re-parameterization} 
\label{sec:ablation-rep}
As Table~\ref{tab:ablation_reparam} showed, we have implemented the re-parameterization technique (rep) \cite{ding2021repvgg} to enhance the efficiency of the inference phase.

\noindent\textbf{Training Setting}
As can be seen from Table \ref{tab:ablation_train}, training the model one more time can improve its performance, but more training rounds will not bring significant gains.

\begin{table}[t]
% \vspace{1em}
% \renewcommand{\arraystretch}{1.2}
    \centering
    % \vspace{-0.5em}
    \scalebox{0.7}{
        \begin{tabular}{c c c c c }
            \toprule
            \multirow{2}{*}{Model} & Set5 & Set14 & BSD100 & Urban100 \\
            ~ & PSNR / SSIM & PSNR / SSIM & PSNR / SSIM & PSNR / SSIM\\
            
            \midrule
            
            SPAN\_no\_rep
            &  32.11/0.8943
            &  28.59/0.7818
            &  27.58/0.7358
            &  26.02/0.7833  \\
            
            SPAN\_rep 
            &  32.18/0.8950
            &  28.63/0.7825
            &  27.60/0.7368
            &  26.10/0.7856  \\
            
            \bottomrule
        \end{tabular}
    }
\vspace{-0.5em}
\caption{Comparision of our baseline SPAN with re-parameterization and SPAN\_no\_rep without using the re-parameterization technology for training process.}
\label{tab:ablation_reparam}
\vspace{-1.5em}
\end{table}

\label{sec:ablation-train}
\begin{table}[th]
% \vspace{1em}
% \renewcommand{\arraystretch}{1.2}
\centering
\vspace{-0.em}    
    \scalebox{0.7}{
        \begin{tabular}{c c c c c }
            \toprule
            \multirow{2}{*}{Model} & Set5 & Set14 & BSD100 & Urban100 \\
            ~ & PSNR / SSIM & PSNR / SSIM & PSNR / SSIM & PSNR / SSIM\\
            
            \midrule
            
            SPAN\_stage1 
            &  32.18/0.8950
            &  28.63/0.7825
            &  27.60/0.7368
            &  26.10/0.7856  \\
            
            SPAN\_stage2 
            &  32.20/0.8953
            &  28.64/0.7828
            &  27.61/0.7370 
            &  26.13/0.7864  \\

            SPAN\_stage3 
            &  32.20/0.8952
            &  28.64/0.7828
            &  27.61/0.7371 
            &  26.13/0.7865  \\
            
            \bottomrule
        \end{tabular}
    }
\caption{Performance comparison of SPAN in different training stages, SPAN\_stage3 represents our latest model.}
\label{tab:ablation_train}
\vspace{-1.5em}
\end{table}

% \begin{table}[!ht]
% \centering
% \resizebox{0.45\textwidth}{!}{
% \begin{tabular}{c|c|cc|cc}
% \hline
%  Act. Type & short-cut & \multicolumn{2}{c}{Run Time (ms)} & \multicolumn{2}{|c}{PSNR (dB)} \\ \hline
% ReLU & yes & x2        & x4       & x2       & x4           \\ \hline
% Sigmoid & yes &         &        &         &              \\ \hline 
% Gelu   & no   &         &        &         &           \\ \hline 
% Leaky ReLU & yes  &         &        &         &                \\ \hline 
% SiLu & no   &         &        &         &          \\ \hline
% \end{tabular}
% }
% \caption{Experiment for \textbf{Guideline 1}. Explore the performance difference using different activation type in each layer to find the best choice.}
% \label{table:guideline_elementwise}
% \end{table}

\vspace{-0.5em}
\subsection{SPAN for NTIRE 2024 challenge}
\vspace{-0.5em}
Our team won the 1st place in the main track (Overall Performance Track) and the 1st place in the sub-track1 (Inference Runtime Track) of NTIRE 2024 efficient super-resolution challenge~\cite{ntire2024efficientsr}. The model structure and training strategy are slightly different from the above. The proposed model has 6 SPABs, in which the number of feature channels is set to 28. DIV2K and LSDIR~\cite{lilsdir} datasets are used for training in this challenge. During each training batch, 64 HR RGB patches are cropped, measuring $256 \times 256$, and subjected to random flipping and rotation. In the training phase, NGswin~\cite{choi2023n} is used as the teacher model for boosting the restoration performance. The learning rate is initialized at $5\times 10^{-4}$ and undergoes a halving process every $2\times10^5$ iterations. The network undergoes training for a total of $10^6$ iterations, with the L1 loss function being minimized through the utilization of the Adam optimizer~\cite{kingma2014adam}. We repeat the aforementioned training settings four times after loading the trained weights. Subsequently, fine-tuning is executed using the L1 and L2 loss functions, with an initial learning rate of $1\times10^{-5}$ for $5\times10^5$ iterations, and HR patch size of 512. We conduct finetuning on four models utilizing both L1 and L2 losses, and employ batch sizes of 64 and 128. Finally, we integrate these four models to obtain the ultimate model. In comparison to the RLFN method that secure the first place in the NTIRE2022 Efficient Super-Resolution Challenge, our method significantly outperforms other methods across all metrics while attaining the fastest running time.

% \subsection{NTIRE 2023 Efficient-SR Challenge}

\begin{table}[t]
% \vspace{1em}
% \renewcommand{\arraystretch}{1.2}
    \centering
    % \vspace{-0.5em}
    \scalebox{0.65}
    {
        \begin{tabular}{c | c c |c  c c c }
            \toprule
            \multirow{2}{*}{Team name} & PSNR & PSNR & Ave Time &
            Parameters & FLOPs   & Overall \\
            ~  & [val] & [test] & [ms] & [M] & [G]  & Ranking \\
            \midrule
            XiaomiMM(ours) & 26.94 & 27.01 & \textbf{5.592} & \textbf{0.151} & \textbf{9.83}  & \textbf{1} \\

            Cao Group & 26.90 &  27.00 & 8.372 & 0.215 & 13.05  & 2 \\

            BSR &  26.90 &  27.00 & 9.384 & 0.218 & 11.95  & 3 \\

            VPEG\_O & 26.90 & 27.01 & 9.630 & 0.212 & 13.86  & 4 \\

            CMVG & 26.90 & 27.01 & 10.022 & 0.202 & 12.17  & 5 \\
            
            % MegSR &  26.95 & 18.30 & 0.243 & 14.90  & 39 \\
    
            % NJUST\_M &  27.05 & 68.11 & 0.104 & 6.56  & 66 \\
            
            % NoahTerminalCV B &  27.03 & 27.83 & 0.203 & 17.7  &64\\
            \midrule
            RLFN(Baseline) & 26.96 & 27.07 & 11.77 & 0.317 & 19.67  & / \\
            \bottomrule
        \end{tabular}
    }
\caption{Mean Results of 5 runs (With an extra zero-run to warm-up the GPU first) on single NVIDIA GeForce RTX 3090 GPU of NTIRE2024 ESR Challenge. The top five methods are included. }
\label{challenge_result}
\vspace{-1.5em}
\end{table}

\vspace{-0.5em}
\section{Conclusion}
\vspace{-0.5em}
In this paper, we have presented the Swift Parameter-free Attention Network, an efficient SISR model that addresses the challenges posed by conventional attention mechanisms in terms of complex network structures, slow inference speed, and large model size. SPAN uses a parameter-free attention mechanism to enhance important information and reduce redundancy. Its simple structure, symmetric activation functions, and residual connections ensure high image quality and fast inference speed. Our extensive experiments on multiple benchmarks have shown that SPAN image quality existing efficient super-resolution models in terms of both performance and inference speed, achieving a significant quality-speed trade-off. This makes SPAN highly suitable for real-world applications, particularly in resource-constrained scenarios such as mobile devices.  Future research may apply the parameter-free attention mechanism to other computer vision tasks and further optimize the network for greater efficiency.

%%%%%%%%% REFERENCES
\clearpage

{\small
\bibliographystyle{ieeenat_fullname}
\bibliography{egbib}
}

\end{document}